\newcommand{\mr}{\mathrm}
\newcommand{\mb}{\mathbf}
\newcommand{\nn}{\nonumber\\[1.5ex]}

\def\co{\; \; ,}

\newcommand{\mpi}{M_{\pi}}
\newcommand{\mpiL}{M_{\pi L}}
\newcommand{\lapi}{\lambda_{\pi}}

\newcommand{\fpi}{F_{\pi}}

\newcommand{\<}{\langle}
\renewcommand{\>}{\rangle}

\newcommand{\cF}{\mathcal{F}}
\newcommand{\ri}{\mr{i}}

\documentclass{PoS}

\usepackage{epsfig,here}

\PoS{PoS(LAT2005)059}

\title{The pion mass in finite volume to two loops }

\ShortTitle{The pion mass in finite volume to two loops }

\author{\speaker{Christoph Haefeli}\\
        Institut f\"ur theoretische Physik\\
        Universit\"at Bern\\
        Sidlerstr. 5, 3012 Bern\\
        E-mail: \email{haefeli@itp.unibe.ch}}

\abstract{We evaluate the pion mass in finite volume to two loops within
  Chiral Perturbation Theory. The results are compared with a recently
  proposed extension of the asymptotic formula of L\"uscher. We find that
  contributions, which were neglected in the latter, are numerically very 
  small at the two-loop level.}

\FullConference{XXIIIrd International Symposium on Lattice Field Theory\\
  25-30 July 2005\\
  Trinity College, Dublin, Ireland}

\begin{document}

\section{Introduction}

\noindent
Numerical simulations with lattice QCD are bound to rather small lattice
volumes when determining the hadron spectrum and other low energy parameters
in QCD. The computed observables show a volume dependence and a thorough
understanding of these effects is important for a correct interpretation of
numerical data. We report on recent progress related to analytical finite
volume studies in case of the pion mass. 

A long time ago L\"uscher established an asymptotic formula
\cite{Luscher:1985dn} which relates the size dependence of the pion mass
$\mpi$ with the $\pi\pi$-forward scattering amplitude $\cF_\pi(\nu)$, 
\begin{equation}
\mpiL-\mpi=-\,{3\over16\pi^2\mpi L}\;
\int_{-\infty}^\infty\!dy\;\cF_\pi(\ri y)\,e^{-\sqrt{\mpi^2+y^2}L}
+O(e^{- \sqrt{3/2}\mpi L})
\co
\label{luscher_mpi_ori}
\end{equation}
with $\mpiL$ the pion mass in finite volume and where the constraint $\mpi
L\gg 1$ is assumed. The contributions in the low energy region are
enhanced, where the $\pi\pi$-scattering amplitude is
represented very accurately through its chiral representation
\cite{Bijnens:1997vq,Colangelo:2001df}. At
leading order in the chiral counting the forward scattering amplitude assumes
to be a constant, and the
integral simplifies to a modified Bessel function of the second kind. When
Colangelo and D\"urr evaluated subleading finite volume effects with
L\"uscher's formula, they observed numerically enhanced corrections with
respect to the leading order \cite{CD}. These investigations show a
clear necessity to go beyond leading order calculations in order to have the
finite volume effects of the pion mass under control. Furthermore, as the
derivation of eq.~(\ref{luscher_mpi_ori}) takes into account only 
exponential contributions of the order $\exp(-\mpi L)$ and systematically
drops those of the order $\exp(-\sqrt{3/2}\mpi L)$, the question
arises, whether the missed terms might turn out to be numerically
relevant. The same question also concerns the asymptotic formulae of decay
constants \cite{Colangelo:2004xr}. As we will show, a full two-loop
calculation of the pion mass in finite volume ChPT clarifies these open
points. 

The two-loop calculation appears to be interesting also for its own right. To
date, a number of finite volume calculations have been performed at one-loop 
order \cite{fvOneloop}, but as far as we know two-loop calculations have only
been performed for the quark condensate \cite{Bijnens:qqbar} and for
low-energy observables in a closely related field \cite{temperature}. As
finite volume effects occur first at the one-loop level, only a two-loop
calculation -- or alternatively if existing, an asymptotic formula \emph{\`a
  la} L\"uscher -- leads to a better understanding of the convergence
behaviour of the perturbative expansion.

\section{ChPT in finite volume}

\noindent
Chiral Perturbation Theory (ChPT) is the effective theory for QCD at low
energies. It is nowadays a mature field which has been applied successfully
for a variety of phenomena, in particular in the meson sector. For an
introduction and a current status of the field, we refer to
Ref.~\cite{ChPTintro,ChPTstatus}.

The effective framework is still appropriate, when the system is enclosed by a
box of size $V=L^3$. We refer to the literature for the foundations
\cite{Gasser:1986vb,Gasser:1987ah,Gasser:1987zq} and a recent review
\cite{Colangelo:2004sc}. Here, we only remind of the fundamental results which
guided the present calculation: the volume has to be large enough, such that
ChPT can give reliable results, $2\fpi L \gg 1$. The perturbative calculation
is bound to the value of the parameter $\mpi L$. Whether it exhibits to be
large ($\mpi L \gg 1$, ''$p$-regime'') or small ($\mpi L \lesssim 1$,
''$\epsilon$-regime'') implies a different power counting. However, in both
cases the effective Lagrangian is the same as in the infinite volume. 
Here, we only cover the ''$p$-regime'', where the system is distorted mildly 
and the only change brought about by the finite volume is a modification of
the pion propagator due to the periodic boundary conditions of the pion fields
\begin{equation}
G(x^0,\mb{x})=\sum_{{\mb n} \in \mathbb{Z}^3}G_0(x^0,{\mb x}+{\mb n}L) \, ,
\label{eq:Gx}
\end{equation}
with $G_0(x)$ the propagator in infinite volume.

\section{Pion in finite volume}

\noindent
The pion mass in finite volume is defined by the pole equation
\begin{equation}
  \label{eq:pole}
  G(\hat{p}_L)^{-1} = 0 \co \qquad \mbox{for} \qquad 
  \hat{p}_L = (i\mpiL,{\bf 0}) \co
\end{equation}
where $G(p^0,{\bf p})^{-1}$ is the Fourier transform of the connected
correlation function
\begin{eqnarray}
  \<\pi^1(x)\pi^1(0)\>_L &=& L^{-3}\sum_{\bf p} \int \frac{dp^0}{2\pi} 
  e^{ipx}\, G(p^0,{\bf p}) \, ,\nn
  G(p^0,{\bf p})^{-1} &=& M^2+p^2-\Sigma_L(p^2) \, ,
\end{eqnarray}
with $\Sigma_L(p^2)$ the self-energy in finite volume and $M^2$ the 
pion mass in the chiral limit in infinite volume. A determination of
the pion mass amounts thus to an evaluation of the self-energy in a loop
expansion. At one-loop order the finite volume corrections have been evaluated
in \cite{Gasser:1986vb}. Here, we discuss the main steps which guided the
two-loop calculation. A detailed derivation of the results will be given
elsewhere \cite{CHmpi2loop}. It is convenient to write the pion mass in finite
volume to two loops in the following manner, 
\begin{eqnarray}
  \label{eq:sum1}
  \mpiL^2 &=& \mpi^2 -\Sigma^{(1)}- \Sigma^{(2)} \co \nn
  \mpi^2  &=& M^2 - \Sigma^{(0)} \co
\end{eqnarray}
where $\Sigma^{(r)}$, $r = 0,1,2$, denote the contribution of the self-energy
with $r$ propagators in finite volume with non-vanishing vector $\bf n$
(cf. eq.~(\ref{eq:Gx}) ). These terms shall be discussed in some detail.

\subsection{Self-energy to 0'th order: $\Sigma^{(0)}$}

\noindent
The contributions to $\Sigma^{(0)}$ are volume independent by definition and
merely renormalize the pion mass, cf. eq.~(\ref{eq:sum1}). A detailed
discussion of this calculation can be found in \cite{Bijnens:1997vq}, with
which we agree.

\subsection{Self-energy to 1'st order: $\Sigma^{(1)}$}

\begin{figure}[t]
\begin{center}
\includegraphics[width=4.5cm]{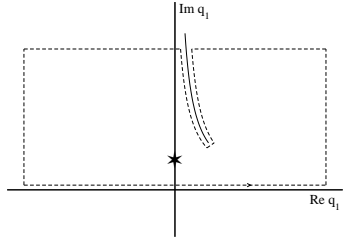}
\end{center}
\caption{Integration contour in the complex $q_1$ plane (dashed line) with the
  pole from the pion propagator (diamond) and the branch cut from the
  $\pi\pi$-scattering amplitude.} 
\label{fig:cut}
\end{figure}

\noindent
As L\"uscher showed \cite{Luscher:1985dn}, the leading finite volume effects
are captured in $\Sigma^{(1)}$ which can be summed up in closed form
\begin{equation}
  \label{eq:sig1}
  \Sigma^{(1)} = \frac{1}{2}\int\frac{d^4q}{(2\pi)^4}
  \sum_{n=1}^{\infty} m(n) \, 
  G_0(q) e^{iq_1\sqrt{n}L}
  \,\Gamma_{\pi\pi}(\hat{p},q,-\hat{p},-q) \co
\end{equation}
with $\hat{p} = (i\mpi,{\bf 0})$, $\Gamma_{\pi\pi}(\hat{p},q,-\hat{p},-q)$
the 4-point function of $\pi\pi$-scattering in the forward scattering
kinematics and $m(n)$ the number of integer vectors ${\bf z}$ with ${\bf
  z}^2=n$. Eq.~(\ref{eq:sig1}) may still be simplified considerably. We
perform a contour integration in the complex plane of the first component
$q_1$. Singularities are met by the pole of the propagator as well as the
branch cuts from the propagator and the 4-point function. At the two-loop
level only the cuts of the 4-point function have to be considered, lying on
a hyperbola in the complex $q_1$ plane, as illustrated in fig.~\ref{fig:cut}. 
Performing the contour integration along the dashed line, we obtain two terms,
one from the residuum of the pole $I_p$ and the other from the
integral along the new integration path. The latter contribution vanishes as
we push the integration lines to infinity, except for the one along
the cut, to be denoted by $I_c$ in the following,
\begin{equation}
  \Sigma^{(1)}   =  I_p + I_c 
  \, .
\end{equation}
 Along the lines discussed in \cite{Luscher:1985dn}, the former
 leads to the resummed asymptotic formula 
\begin{equation}
  \label{eq:mod_luscher}
  I_p =
  {\mpi^2\over16\pi^2\lapi}\;\sum_{n=1}^{\infty}{m(n)\over\sqrt{n}} 
  \int\limits_{-\infty}^\infty\!dy\;
  \cF_\pi(\ri y)\,e^{-\sqrt{n(1+y^2)}\lapi} \co 
\end{equation}
with $\cF_\pi(iy)$ the $\pi\pi$-forward-scattering amplitude.
Restricting the sum to the first
addend, we recover L\"uscher's formula, eq.~(\ref{luscher_mpi_ori}). Its
extension to eq.~(\ref{eq:mod_luscher}) has already been suggested in
\cite{Colangelo:2004sc,Colangelo:2005gd}. Numerically, the resummation
turns out to be relevant for moderate $\lapi$.  

Concerning the contributions of the cut $I_c$, we make use of a dispersive
treatment. As these terms only start at the two-loop level, they are both
suppressed in the chiral as well as in the large $L$ expansion and are
therefore expected to be small. This is exactly what we observe numerically.

\subsection{Self-energy to 2'nd order: $\Sigma^{(2)}$}

\noindent
Contributions from two pion propagators in finite volume are finally captured
in $\Sigma^{(2)}$. Notice that the corresponding Feynman diagrams are
uv-finite and need not to be renormalized. Therefore, it only remains to find
a feasible numerical representation for these terms.

\section{Numerics}
\label{sect:numerics}

\begin{figure}[t]
\begin{center}
\includegraphics[width=8cm]{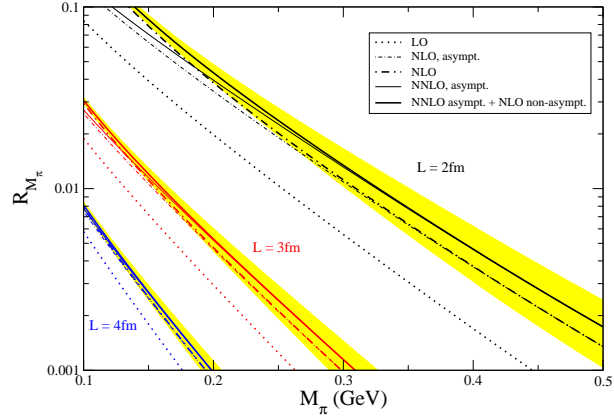}
\end{center}
\caption{$R_{\mpi}=\mpiL/\mpi-1$ vs. $\mpi$ for $L=2,3,4$fm. For explanations
  of the legend, see text.}
\label{fig:Rmpi}
\end{figure}

\noindent
The numerical analysis is performed in line with the setup of
Ref.\cite{Colangelo:2005gd}. In fig.~\ref{fig:Rmpi}, we evaluate the relative 
finite volume shift 
\begin{equation}
  R_{\mpi} \equiv \frac{\mpiL-\mpi}{\mpi} \co
\end{equation}
for $L=2,3,4$ fm as a function of $\mpi$. We show the one-loop (LO) as
well as the two-loop result (NLO). These shall be compared with the resummed
asymptotic formula eq.~(\ref{eq:mod_luscher}) with LO/NLO/ NNLO input for the
$\pi\pi$-scattering amplitude. Note that the one-loop result and the resummed
asymptotic formula to LO agree with each other. The best estimate for
$R_{\mpi}$ is finally obtained by adding to the asymptotic pure three-loop
contribution the two-loop result (NNLO asympt. + NLO non-asympt.). At NLO, the
finite size effects encounter low energy constants, leading to a
non-negligible error band which is only shown for the best estimate. 
We take up a point already alluded in the introduction, namely the large
contributions when going from LO to NLO in the asymptotic formula (dotted to
thin-dash-dotted). Compared to this gap, the additional contributions from the
full two-loop result (thick-dash-dotted) are very small. Consider eg. a pion
mass of $\mpi=250$ MeV in a 2 fm box. We find $R_{\mpi}=0.0236(41)$, of which
0.0010 stem from the two-loop corrections which are not included in the
asymptotic formula. The two-loop and the NLO result from the
asymptotic formula only drift away, when we go beyond the region where the
$p$-regime can be safely applied.  

\section{Conclusions}

\noindent
We have evaluated the finite volume effects of the pion mass to two loops
within ChPT in the $p$-regime. The results are compared with a recently
proposed extension of the asymptotic formula of L\"uscher. We find that
contributions which were neglected in the latter, are numerically very small
at the two-loop level and conclude that the resummed asymptotic formula is a
convenient method to evaluate the finite volume effects beyond the
leading order.

\section*{Acknowledgements}

\noindent
The work presented here is being done in collaboration with Gilberto Colangelo
whom I warmly thank, also for a careful reading of the manuscript. 
This work is supported by the Swiss National Science Foundation
and in part by RTN, BBW-Contract No. 01.0357 and EC-Contract
HPRN--CT2002--00311 (EURIDICE).

\end{document}